# Earthquake-Like Avalanches in Compression of Ductile Porous Materials

Hao Lin[1,2], HaiYing Wang[1,2], Chunsheng Lu[3] and LanHong Dai[1,2,*]

[1]*State Key Laboratory of Nonlinear Mechanics, Institute of Mechanics, Chinese Academy of Sciences, Beijing 100190, China*

[2]*School of Engineering Science, University of Chinese Academy of Sciences, Beijing 101408, China*

[3]*School of Civil and Mechanical Engineering, Curtin University, Perth, Western Australia 6845, Australia*

As a complex natural phenomenon, seismicity can be characterized by several power laws. In the viewpoint of brittle fracture in disordered materials, however, these scaling behaviors, especially their critical exponents, are still far from understanding. In this Letter, avalanches in compression of ductile metallic glass foam are investigated, in which inelasticity can be naturally introduced by strain localization (i.e., nanoscale shear bands) prior to fracture of cellular structures. It is shown that avalanche statistics in such a process can reproduce fundamental seismic power laws with exponents compatible with earthquakes. It implies that in addition to elasticity and disorder involved in conventional statistical models, the inelastic rheological behavior is a necessary ingredient in accounting for seismological laws.

Earthquakes are believed as one of the most complex natural phenomena, which can be characterized by several empirical power laws: (i) The Gutenberg-Richter (GR) law states that the number of earthquakes as a function of their size or energy $S$ decreases as a power law, i.e., $P(S)dS \sim S^{-\tau}dS$ (with $\tau = 1 + 2b/3$ and $b \approx 1$) [1]. (ii) The rescaled time interval between two successive earthquakes (or waiting times) can be described by a universal gamma distribution, $P(x) \sim x^{\gamma-1}\exp(-x/b)$, in regimes with stationary seismic activities [2]. (iii) The Omori law indicts that the rate of aftershocks decreases with the time $t$ after a main shock as $1/t^p$ (with $p = 1$ for large earthquakes) [3]. Finally, (iv) the productivity law gives the number of aftershocks, which are trigged by the event size $S$, scales as $S^{2\alpha/3}$ (with $\alpha \approx 0.8$) [4]. Here, how to represent these seismic laws in experiments or simulations with minimal physics is of great significance in understanding an earthquake generation process.

Recently, some statistical similarities have been found between earthquakes and acoustic emission signals recorded during compression of brittle porous materials [5]. That is, in terms of crackling noise, fracture of disordered materials is related to earthquakes [6], and in both systems, slow perturbation responds through discrete and scale-free events with a huge variety of sizes. It implies the existence of a criticality that has stimulated testing on the universality of power-law exponents from various brittle heterogeneous or porous media [7−10]. Although a number of efforts have been made, there is still lack of an understanding on these seismic laws. First, experimentally accessible quantities in brittle fracture (BF) such as the energy or avalanche size distributions are not straightforwardly comparable with those measured in earthquakes. The GR law exponents ($\tau =$ 1.4 ± 0.1 [5, 7−10]) observed in BF are obviously smaller than that in earthquakes ($\tau \approx 1.67$ [1]).

The similar phenomenon exists in conventional statistical models, such as the Olami-Feder-Christensen [11] and elastic depinning models [12]. Without considering the rheological properties of materials, these elastic-brittle models usually give an upper limit value of the avalanche size exponent, i.e., $\tau = 1.5$ [13] obtained by the mean field theory, which is also smaller than $\tau \approx 1.67$. In addition, based on available BF experiments, the distributions of waiting time exhibit non-stationary characteristics of a double power law [5–10, 14], significantly different from the unified scaling behavior of waiting time for earthquakes [2]. All these imply that simple BF fails to reveal information hidden in the seismic laws.

Generally speaking, earthquakes occur due to the instability of a fault system filled with crushed pieces of rocks (fault gouge [15]). Analogous to amorphous materials like metallic glasses [16], the irreversible deformation of granular fault gouge is usually through rearrangements of a small cluster of particles and spontaneous localization into narrow shear bands at a critical external stress [15]. This may lead to severe temperature rise and significant rheological change in local materials. Considering such an inelastic effect as additional viscoelastic relaxation in modified stick-slip models, novel avalanche dynamics were observed with quasi-periodic stress oscillations characterized by a higher avalanche size exponent [17]. These findings shed light on a new mechanism to account for seismological laws, however, a direct experimental evidence for the inelastic effect with seismic activities is still lacking. On the other hand, simulations can provide a detailed description of viscoelastic dynamics in disordered systems at a short timescale [18], but their long-term clustering, scaling and universality remain poorly understood. In this Letter, we deal with the ductile metallic glass foam (DMGF), a typical kind of porous materials. Different from its brittle counterpart, DMGF can release local excessive stress by forming nanoscale shear bands in cellular structures before fracture [19, 20], which provides a good physical approximation of plastic strain localization in fault gouge layers. In contrast to a simple BF process, we demonstrate that, by naturally introducing the inelastic effect via shear-banding interaction, the system can generate self-similar avalanches that represent all fundamental seismic laws.

*Experiments.*— Cylindrical metallic glass foams with 10 mm in diameter and porosity $\phi = 60\ \%$ were fabricated by pressure infiltration combined with rapid cooling [20]. For comparison, brittle alumina foam (BAF) with 30 mm in diameter and porosity $\phi = 87\ \%$ were also prepared. Uniaxial compression tests were carried out on the MTS-810 material test system at ambient temperature. The displacement-driven mode [21] was adopted with a constant initial strain-rate of $10^{-3}$ /s for all samples. The force opposed by the sample was measured with a high resolution where the nominal uncertainties are of the order 0.1 N. The sampling frequency was 100 HZ and it is well enough to capture each peak or critical stress which separates the transition of deformation from pinning to burst slip. As shown in Fig. 1(a), deformation of BAF proceeds by brittle fracture of disordered cellular structures along the main crack tip, in which elasticity and disorder plays an important role in the process of energy balance; while in DMGF, plastic strain is locally accumulated in foam's cellular walls by formation of nanoscale shear bands prior to fracture [19, 20]. This is also reflected from the stationarity of their force-time curves as shown in Figs. 1(b) and (c): for BAF, force varies significantly over time, i.e., randomly transient hardening or softening, however for DMGF, there are obviously stationary oscillations, where pronounced fluctuations of the mean stress in each serration, i.e., a loading and unloading cycle, are absent. In order to quantify such a difference, let us introduce the avalanche, *S*, following Ref. [22], and consider the sequence of external force *F*(*t*), as shown in Fig. 1(d), where bursts appear as 'valleys'. It is of interest to note that there is a hierarchical structure in large valleys, inside which are some smaller ones. Here we define a burst corresponding to time $t_i$ as the total irreversible deformation increment in the forward direction, i.e., $S_i = \sum_{j=i}^{iend} \Delta x_j$, in the interval starting from the current yield stress, $F_{max,i}$, to a larger one, $F_{max,iend+1}$, where the local burst displacement $\Delta x_j$ is calculated by

$$\Delta x_j = \frac{F_{max,j}}{k_{j-1}} - \frac{F_{min,j}}{k_j}, \quad (1)$$

where $F_{max,j}$ (or $F_{min,j}$) and $k_j$ are the local maximum (or minimum) force and stiffness at time $t_j$, respectively. With the time evolution, we can obtain an avalanche series that is considered to be an earthquake-like sequence. Fig. 1(e) shows the typical time evolution of avalanche events (the size smaller than 100 nm are neglected) and their cumulative number (counted at an interval of 0.1 s). Each sample produces typically a few thousand events: 1700 ± 500 and 16000 ± 1000 for DMGF and BAF, respectively.

*The GR law exponent* (1.5 *versus* 1.7).— To obtain the maximum resolution of a limited experimental data set, the complementary cumulative distribution function (CCDF) was used. As shown in Figs. 2(a) and (b), the avalanche size distributions follow the GR law, i.e., CCDF(*S*) ~ $S^{-(\tau-1)}$, over several magnitudes for both BAF and DMGF, respectively, and thus their corresponding probability density functions scale as $P(S) \sim S^{-\tau}$. In the case of BAF, $P(S)$ is well fitted by using the maximum likelihood method, with a robust power law exponent $\tau = 1.43 \pm 0.05$, which is quite close to $\tau = 1.4 \pm 0.1$ obtained by acoustic emission signals in various brittle porous materials [5–10] and $\tau$ = 1.5 predicted by the elastic models in the mean field limit [13]. However, all of them are clearly smaller than $\tau$ = 1.67 observed in earthquakes [23]. While in DMGF, we found a significantly different scaling exponent, $\tau$ = 1.70 ± 0.05, which is more comparable with earthquakes. A larger GR law exponent means that the extreme events in DMGF decay faster than those in BAF. This can be attributed to additional plastic shear bands in DMGF, which effectively prevented the non-steady propagation of cracks and fragmented an intact avalanche into several smaller ones. The similar fragmentation effect on an avalanche size distribution was also discussed by using a forest fire model [24].

*The waiting time* (*a double power law versus the Gamma distribution*).— Another important aspect of earthquakes is related to the waiting time, $\tau$, between two successive events above a lower bound avalanche size $S_m$. Figs. 3(a) and (c) show the probability density functions, $P(\tau)$, obtained from BAF and DMGF, respectively, with several values of $S_m$. Bak *et al*. [25] proposed that after accounting for the spatial location of events, the distributions of $\tau$ collapse onto a single curve without distinguishing among foreshocks, main shocks, and aftershocks. Corral [2] argued that the occurrence of earthquakes differs regionally and suggested an extension held for different spatial areas, time windows, and magnitude ranges by introducing a mean rate of seismic activities $<r>$ or $1/<\tau>$ in the scaling operation. Then, a more generalized scaling formula is given by [26]

$$P(\tau) = \langle r \rangle^{\omega/(2-\omega)} F\left(\tau \langle r \rangle^{1/(2-\omega)}\right), \quad (2)$$

where $\omega$ and $F(x)$ are the universal scaling exponent and function, respectively. For earthquakes, $\omega \approx 1$ and $F(x)$ can be adjusted by a gamma distribution, $F(x) \sim x^{-(1-\gamma)}\exp(-x/b)$ with $\gamma$ and $b$ the fitting parameters [2]. However, in BAF, there is an obviously different scaling exponent $\omega \approx 1.5$ and a power law decaying function $F(x) \sim x^{-\omega}$ in the main time range [see Fig. 3(b)], which implies the temporal correlation in deformation dynamics. Moreover, a second power law emerges for the rightmost tail of distributions, with an exponent of 3. It leads to a double power-law master curve as proposed in Ref. [2] for the system with a nonstationary activity rate, which is analogous to other brittle porous materials [7, 8]. On the other hand, in DMGF, the scaling exponent $\omega$ is close to 1. Thus, according to Eq. (2), $P(\tau)$ can be rescaled by the mean rate of activities $<r>$. It is shown that there is a good collapse of all distributions [see Fig. 3(d)] by using a universal scaling function, $F(x) \sim x^{-(1-\gamma)}\exp(-x/b)$ with $\gamma = 0.77 \pm 0.06$ and $b = 1.4 \pm 0.1$. This is statistically indistinguishable from the results of earthquake data given in Ref. [2] ($\gamma = 0.67 \pm 0.05$ and $b = 1.58 \pm 0.15$).

*Aftershocks* (*the Omori and productivity laws*).—We further considered main shocks as all the events with the size larger than $S_m$. After a main shock, the sequence of subsequent events was studied until an event with the size larger than that of the main shock. Then, we divided the time from a main shock toward future into intervals, for which the number of aftershocks was counted in each of them. Averages of different sequences corresponding to main shocks with $S > S_m$ were performed. As shown in Figs. 4(a) and (c), the number of aftershocks per unit of time, $r_{AS}(t - t_{MS})$, decays as a power-law function, $r_{AS} \sim 1/(t - t_{MS})^p$, where $t - t_{MS}$ is the elapsed time since the main shock. This relationship hold for brittle and ductile heterogeneous materials with different microscopic mechanisms, however, which gives rise to a significant difference of $p$-values. For BAF where the transient stress divergences come from BF in local cellular structures or rate-and-state-dependent friction between crack surfaces [27], the Omori decay is observed up to three decades of magnitude with $p = 0.5 \pm 0.1$, which is relatively smaller than that in earthquakes, where $p$ usually lies in the range of $0.7 - 1.5$ [3]. For DMGF, the criticality in temporal correlations can be prevented by viscoelastic dissipation [28], resulting in a faster decaying rate of aftershocks with a higher $p$-value ($0.80 \pm 0.05$), which is consistent with observations in earthquakes. Furthermore, by rescaling the vertical axis with $S_m^{-2\alpha/3}$ and the optimum $\alpha$ that leads to the collapse of the data, $r_{AS}/S_m^{2\alpha/3}$ should be a function of the time since the main shock. As shown in Figs. 4(b) and (d), different from other scaling exponents as discussed, $\alpha$ exhibits a clear insensitivity of material rheological behaviors and almost shares the same value for brittle and ductile porous materials ($\alpha \approx 0.66$ for BAF and $\alpha \approx 0.70$ for DMGF), which are both comparable with ones reported in earthquakes ($\alpha \in 0.7$–$0.9$ [4]).

*Conclusion.*— We have presented a complete comparison of the avalanche statistics (including the distributions of avalanche size and waiting time as well as the modified Omori's law) between two types (brittle and ductile) of porous materials and earthquakes. It is shown that in contrast to BF, the additional shear-banding interaction in DMGF, which introduces complex internal dynamics via viscoelastic relaxation, not only leads to distinct avalanche size distributions characterized by a higher power-law exponent, but also induces tempo-spatial correlated aftershock sequences and unique long-term clustering characterized by a universal rescaled waiting time distribution, which are more comparable with earthquakes. Our findings demonstrate that the inelastic rheological behavior is a necessary ingredient in accounting for fundamental seismological laws, which cannot be ignored as assumed in conventional earthquake models. Further studies are needed to advance the models that can reproduce the universal rescaled waiting time distribution and a tempo-spatial correlated aftershock sequence.

This work was supported by the National Key Research and Development Program of China (No. 2017YFB0702003), the National Natural Science Foundation of China (Nos. 11672316 and 11790292), and the Opening Fund of State Key Laboratory of Nonlinear Mechanics. C. Lu is also grateful to the support from the Open Fund of State Key Laboratory for GeoMechanics and Deep Underground Engineering, China University of Mining & Technology (Beijing) (SKLGDUEK1516).

*lhdai@lnm.imech.ac.cn

[1] B. Gutenberg and C. F. Richter, Bull. Seismol. Soc. Am. **46**, 105 (1956); L. Knopoff, Y. Y. Kagan, and R. Knopoff, Bull. Seismol. Soc. Am. **72**, 1663 (1982); T. Utsu, Pure Appl. Geophys. **155**, 509 (1999).

[2] A. Corral, Phys. Rev. E **68**, 035102 (2003); Phys. Rev. Lett. **92**, 108501 (2004); Physica A **340**, 590 (2004).

[3] T. Utsu, Y. Ogata, and S. Matsu'ura, J. Phys. Earth. **43**, 1 (1995); R. Shcherbakov, D. L. Turcotte, and J. B. Rundle, Geophys. Res. Lett. **31**, 11613 (2004); D. Sornette and G. Ouillon, Phys. Rev. Lett. **94**, 038501 (2005).

[4] A. Helmstetter, Phys. Rev. Lett. **91**, 058501 (2003).

[5] J. Baró, A. Corral, X. Illa, A. Planes, E. K. H. Salje, W. Schranz, D. E. Soto-Parra, and E. Vives, Phys. Rev. Lett. **110**, 088702 (2013).

[6] J. P. Sethna, K. A. Dahmen, and C. R. Myers, Nature **410**, 242 (2001).

[7] G. F. Nataf, P. O. Castillo-Villa, J. Baró, X. Illa, E. Vives, A. Planes, and E. K. H. Salje, Phys. Rev. E **90**, 022405 (2014).

[8] T. Makinen, A. Miksic, M. Ovaska, and M. J. Alava, Phys. Rev. Lett. **115**, 055501 (2015).

[9] S.-T. Tsai, L.-M. Wang, P. Huang, Z. Yang, C.-D. Chang, and T.-M. Hong, Phys. Rev. Lett. **116**, 035501 (2016).

[10] H. V. Ribeiro, L. S. Costa, L. G. A. Alves, P. A. Santoro, S. Picoli, E. K. Lenzi, and R. S. Mendes, Phys. Rev. Lett. **115**, 025503 (2015).

[11] Z. Olami, H. J. S. Feder, and K. Christensen, Phys. Rev. Lett. **68**, 1244 (1992).

[12] D. S. Fisher, Phys. Rep. 301, 113 (1998); M. Kardar, Phys. Rep. **301**, 85 (1998).

[13] S. Zapperi, K. Bækgaard Lauritsen, and H. E. Stanley, Phys. Rev. Lett. **75**, 4071 (1995); K. A. Dahmen, Y. Ben-Zion, and J. T. Uhl, Phys. Rev. Lett. **102**, 175501 (2009).

[14] S. Hainzl, G. Zoller, and J. Kurths, J. Geophys. Res. **104**, 7243 (1999); E. A. Jagla and A. B. Kolton, J. Geophys. Res. **115**, B05312 (2010).

[15] F. M. Chester, and J. S. Chester, Tectonophysics, **295**, 199 (1998); N. D. Paola, C. Collettini, D. R. Faulkner, and F. Trippetta, Tectonics, **27**, 4017 (2008).

[16] F. Spaepen, Acta Metall. **25**, 407 (1977); A. S. Argon, Acta Metall. **27**, 47 (1979).

[17] S. Papanikolaou, Phys. Rev. E **93**, 032610 (2016); L. E. Aragón, E. A. Jagla and A. Rosso, Phys. Rev. E **85**, 046112 (2012).

[18] E. A. Jagla, F. P. Landes and A. Rosso, Phys. Rev. Lett. **112**, 174301 (2014).

[19] M. D. Demetriou, C. Veazey, J. S. Harmon, J. P. Schramm, and W. L. Johnson, Phys. Rev. Lett. **101**, 145702 (2008).

[20] H. Lin, H. Y. Wang, C. Lu, and L. H. Dai, Scr. Mater. **119**, 47 (2016).

[21] K. M. Larson, J. T. Freymueller, and S. Philipsen, J. Geophys. Res. Solid Earth **102**, 9961 (1997).

[22] S. Roux and E. Guyon, J. Phys. A **22**, 3693 (1989); A. Tanguy, M. Gounelle, and S. Roux, Phys. Rev. E **58**, 1577 (1998).

[23] L. Knopoff, Y. Y. Kagan, and R. Knopoff, Bull. Seismol. Soc. Am. **72**, 1663 (1982); T. Utsu, Pure Appl. Geophys. **155**, 509 (1999); Y. Y. Kagan, Tectonophysics **490**, 103 (2010); C. Godano, E. Lippiello, and L. de Arcangelis, Geophys. J. Int. **199**, 1765 (2014).

[24] E. A. Jagla, Phys. Rev. Lett. **111**, 238501 (2013).

[25] P. Bak, K. Christensen, L. Danon, and T. Scanlon, Phys. Rev. Lett. **88**, 178501 (2002).

[26] J. Barés, PhD Thesis, Ecole Polytechnique, 2013.

[27] A. Ruina, J. Geophys. Res. Solid Earth **88**, 10359 (1983).

[28] S. Hainzl, G. Zöller, and J. Kurths, J. Geophys. Res. Solid Earth **104**, 7243 (1999).

**Figure captions**

**FIG. 1** (Color online). **(a)** Microscopic deformation of the two types of porous materials: in BAF, deformation proceeds by random fracture of disordered cellular structures along the main crack tip; while in DMGF, deformation is through competition between crack growth and local plastic slips, i.e., nanoscale plastic shear bands. **(b)** and **(c)** show the typical force time curves (with a constant strain rate of $1 \times 10^{-3}$ s$^{-1}$) for BAF and DMGF, respectively. **(d)** The avalanche size, $S$, can be easily determined from the experimental force time curves [19]. **(e)** The typical time evolution of avalanche size and the total number of events in DMGF.

**FIG. 2** (Color online). **(a)** and **(b)** show the avalanche size distributions that follow the GR law, i.e., CCDF($S$) ~ $S^{-(\tau-1)}$, and the corresponding probability density function $P(S)$ ~ $S^{-\tau}$, over several decades of magnitude, respectively. For BAF, $P(S)$ is well fitted by using the maximum likelihood method with a robust power law exponent $\tau = 1.43 \pm 0.05$, while for DMGF, $\tau = 1.70 \pm 0.05$, which is more comparable with $\tau = 1.67$ observed in earthquakes.

**FIG. 3** (Color online). **(a)** and **(c)** show the distributions of waiting times with $S > S_m$ for BAF and DMGF, respectively. Each curve is associated with a value of $S_m$. **(c)** and **(d)** are the waiting times and their distributions rescaled by $<r>^{1/(2-\omega)}$ and $<r>^{\omega/(2-\omega)}$, respectively, where $<r>$ is the mean rate of activities. The values of $\omega$ are significantly different in BAF ($\omega \approx 1.5$) and DMGF ($\omega \approx 1$). The red circles for BAF and blue ones for DMGF are the window average over all distributions. Moreover, the unified scaling functions for the two types of porous materials are also different: a double power-law for BAF and the gamma distribution for DMGF.

**FIG. 4** (Color online). **(a)** and **(c)** Number of aftershocks per unit time, $r_{AS}$, as a function of the time distance to a main shock, $t - t_{MS}$, for BAF and DMGF, respectively, where main shocks are defined as the events with size $S > S_m$ and each curve is associated with a value of $S_m$. The dashed lines are power-law functions scaled as $(t - t_{MS})^{-p}$, which indicate different $p$ values for BAF ($p = 0.5 \pm 0.1$) and DMGF ($p = 0.80 \pm 0.05$). **(b)** and **(d)** Rescaled Omori plot shows the productivity law, with $\alpha \approx 0.66$ for BAF and $\alpha \approx 0.7$ for DMGF.

**FIG. 1**

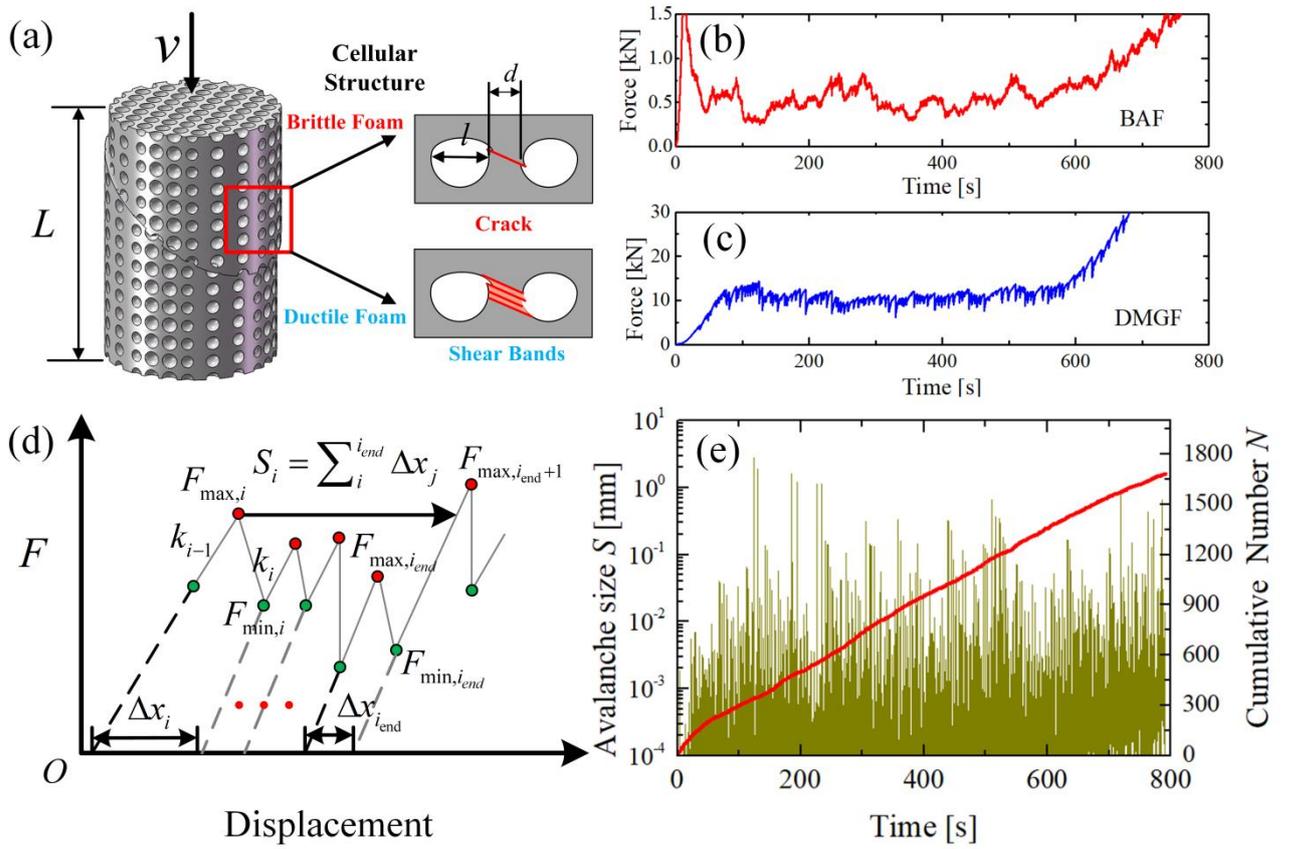

**FIG. 2**

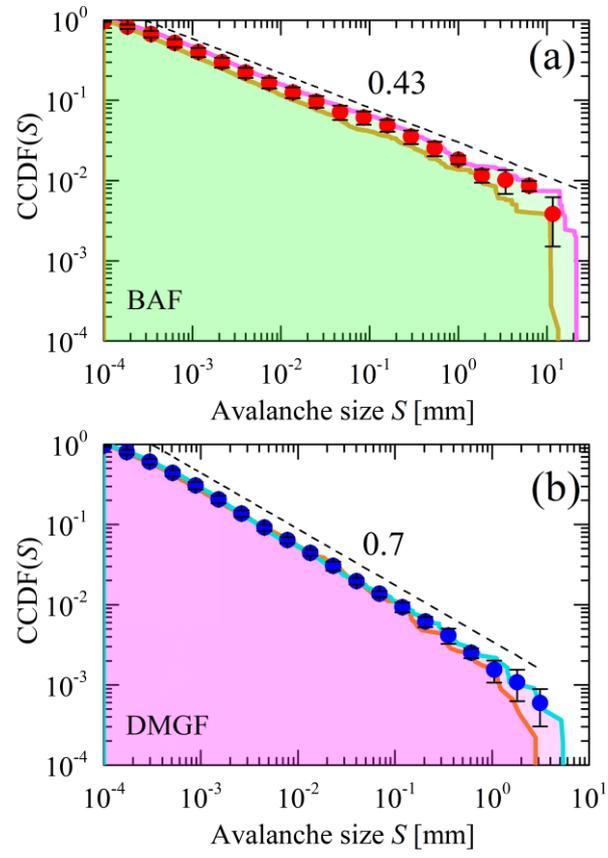

**FIG. 3**

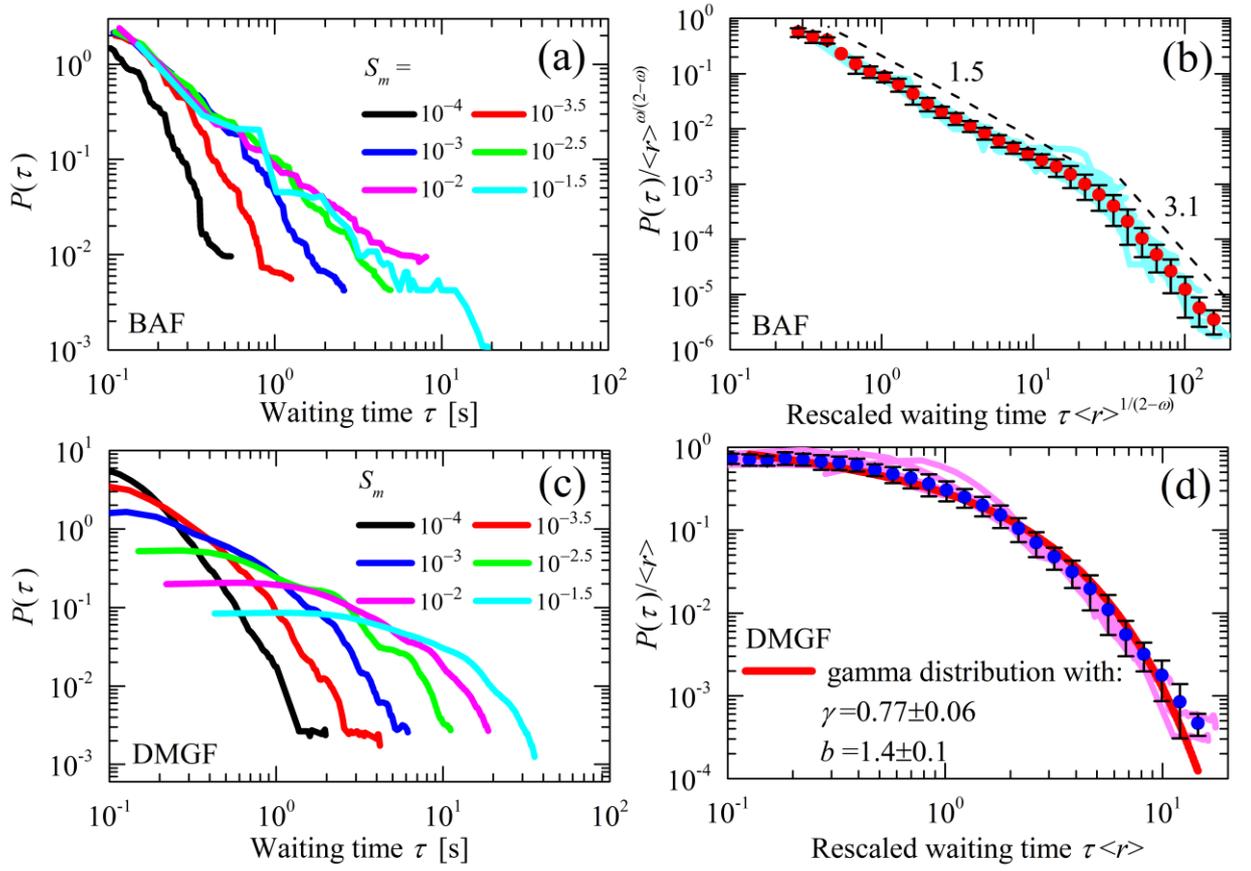

**FIG. 4**

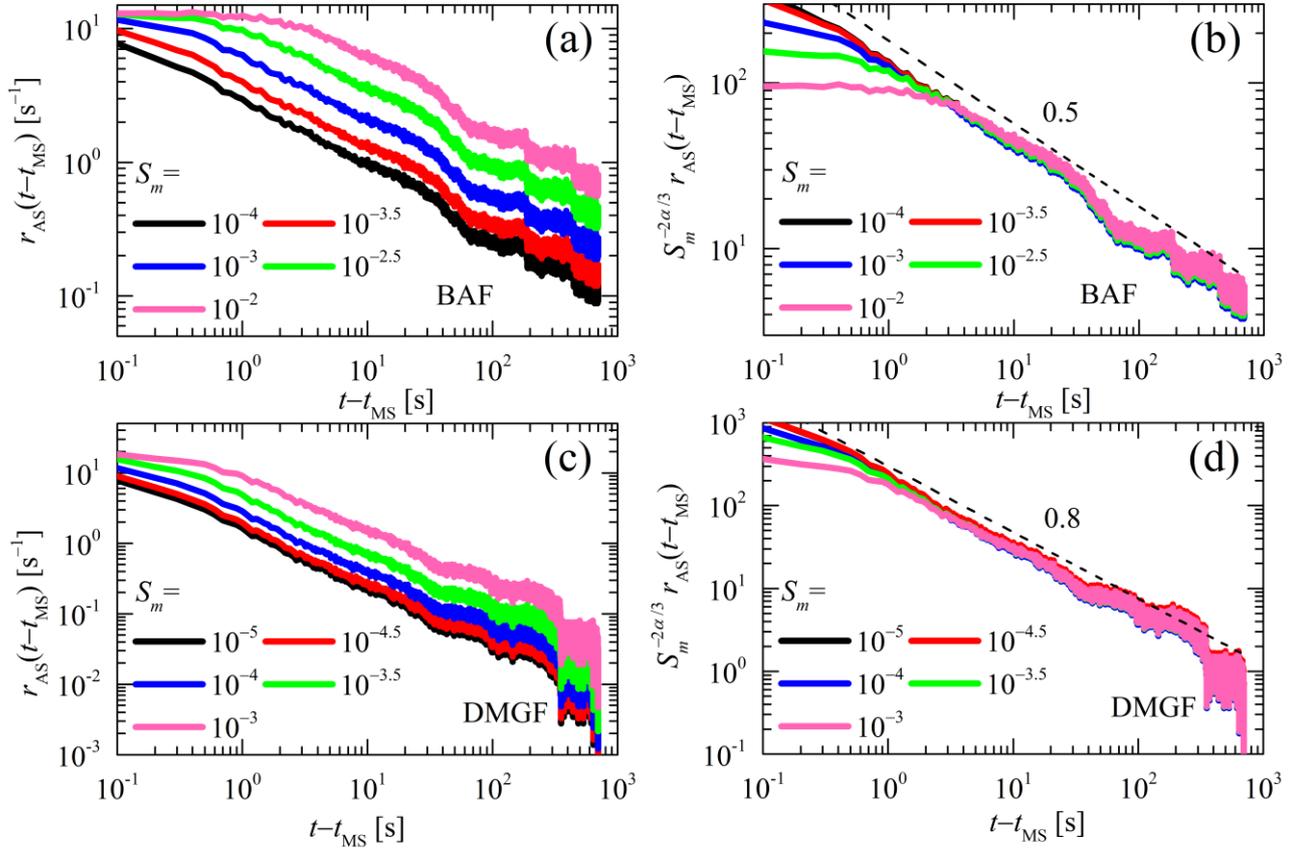